\documentclass[conference]{IEEEtran}
\IEEEoverridecommandlockouts
\usepackage{cite}
\usepackage{amsmath,amssymb,amsfonts}
\usepackage{algorithmic}
\usepackage{graphicx}
\usepackage{textcomp}
\usepackage{xcolor}
\usepackage[compatibility=false]{caption}
\usepackage{subcaption}
\usepackage{adjustbox}
\usepackage{tabularx}

\def\BibTeX{{\rm B\kern-.05em{\sc i\kern-.025em b}\kern-.08em
    T\kern-.1667em\lower.7ex\hbox{E}\kern-.125emX}}

\begin{document}

\title{Data-Driven Assessment of Concrete Slab Integrity via Impact-Echo Signals and Neural Networks\\
}

\author{
% \IEEEauthorblockN{Anonymous submission}

\IEEEauthorblockN{Yeswanth Ravichandran$^*$}
\IEEEauthorblockA{\textit{College of Engineering and Computing} \\
\textit{George Mason University}\\
Fairfax, USA \\
yravicha@gmu.edu}
\and
\IEEEauthorblockN{Duoduo Liao$^*$}
\IEEEauthorblockA{\textit{School of Computing} \\
\textit{George Mason University}\\
Fairfax, USA \\
dliao2@gmu.edu}
\and
\IEEEauthorblockN{Charan Teja Kurakula$^*$}
\IEEEauthorblockA{\textit{College of Engineering and Computing} \\
\textit{George Mason University}\\
Fairfax, USA \\
ckurakul@gmu.edu}

}

%%%% -- add a short line for footnote
\renewcommand{\footnoterule}{%
  \kern -3pt
  \hrule width 0.2\textwidth
  \kern 2.6pt
}

\maketitle

\def\thefootnote{*}\footnotetext{Equal contribution.}

\begin{abstract}
Subsurface defects such as delamination, voids, and honeycombing critically affect the durability of concrete bridge decks but are difficult to detect reliably using visual inspection or manual sounding. This paper presents a machine learning based Impact Echo (IE) framework that automates both defect localization and multi-class classification of common concrete defects. Raw IE signals from Federal Highway Administration (FHWA) laboratory slabs and in-service bridge decks are transformed via Fast Fourier Transform (FFT) into dominant peak-frequency features and interpolated into spatial maps for defect zone visualization. Unsupervised k-means clustering highlights low-frequency, defect-prone regions, while Ground Truth Masks (GTMs) derived from seeded lab defects are used to validate spatial accuracy and generate high-confidence training labels. From these validated regions, spatially ordered peak-frequency sequences are constructed and fed into a stacked Long Short-Term Memory (LSTM) network that classifies four defect types shallow delamination, deep delamination, voids, and honeycombing with 73\% overall accuracy. Field validation on the bridge deck demonstrates that models trained on laboratory data generalize under realistic coupling, noise, and environmental variability. The proposed framework enhances the objectivity, scalability, and repeatability of Non-Destructive Evaluation (NDE), supporting intelligent, data-driven bridge health monitoring at a network scale.
\end{abstract}


% ---------------- Introduction (UNCHANGED) ----------------
\section{Introduction}
\label{sec: intro}
\subsection{Background and Motivation}
Concrete infrastructure, including bridges, buildings, dams, and transportation networks, constitutes a critical foundation for economic activity and public safety, but continues to deteriorate over time due to delamination, honeycombing, voids, and cracking. In the United States, the 2021 ASCE Report Card estimates that approximately 7.5\% of 617,000 bridges are structurally deficient, 42\% are over 50 years old, and the repair backlog stands at around \$125 billion. Addressing these deficiencies requires increasing annual spending from \$14.4 billion to \$22.7 billion, a 58\% rise~\cite{ASCE2021}. Subsurface defects pose particular challenges because they develop beneath apparently intact surfaces, escape visual inspection, and accelerate deterioration through moisture ingress and corrosion. Traditional manual inspection methods, such as chain-drag and sounding, remain limited by practitioner variability and subjectivity~\cite{Carino2013}. To overcome these challenges, Non-Destructive Evaluation (NDE) methods, especially the Impact Echo (IE) technique, are widely adopted to infer concrete thickness and detect delamination from resonance peaks~\cite{Sansalone1989, Sansalone1997, ASTM2015, Carino1986}. IE-based diagnosis relies primarily on frequency-domain signal analysis using the Fast Fourier Transform (FFT) to extract dominant resonance frequencies~\cite{cochran1967fft}, complemented by time frequency and Empirical Mode Decomposition (EMD) variants to enhance robustness against noise and modal interference \cite{Zhang2012, Melhem2003}. Recent research increasingly integrates machine learning with IE data to improve scalability and reduce human subjectivity in defect detection. Both shallow and deep learning models are being applied, ranging from Convolutional Neural Networks (CNNs) for image-based surface crack detection \cite{Cha2017, Dorafshan2018} to sequence models such as Recurrent Neural Networks (RNNs), including Long Short-Term Memory (LSTM) networks for learning spectral and temporal patterns \cite{Hochreiter1997}. Advanced studies further incorporate physics-guided labeling, multimodal fusion, and automated contour interpretation to enhance bridge deck assessment, improve accuracy, and facilitate the real-world deployment of NDE systems \cite{rachuri2024, pavurala2024, dorafshan2020, darji2024,sengupta2023}.

\subsection{Research Gap and Problem Statement}

Despite the proven effectiveness of IE testing for concrete assessment, current methodologies predominantly rely on manual interpretation of frequency spectra and human expertise for defect identification and classification. These approaches introduce several critical limitations: 

1) \textit{Subjectivity and inconsistency}: Frequency response interpretation varies among analysts, leading to non-reproducible assessments and dependency on operator experience. 

2) \textit{Temporal inefficiency}: Manual analysis of large datasets from extensive bridge networks requires prohibitive time investments, limiting inspection frequency.

3) \textit{Limited scalability}: Specialized operator training requirements constrain inspection capacity across bridge networks.

4) \textit{Binary classification}: Conventional approaches typically distinguish only between ``defective'' and ``non-defective'' regions without characterizing defect type, depth, or severity. 

5) \textit{Spatial context neglect}: Point-by-point analysis often overlooks spatial patterns and regional trends that inform defect characterization.

\noindent While recent studies have explored signal processing enhancements for IE data using techniques such as wavelet decomposition~\cite{Melhem2003} and EMD~\cite{Zhang2012}, and applied basic machine learning for defect detection~\cite{Liao2016}, comprehensive frameworks that integrate spatial analysis, unsupervised anomaly detection, and deep learning architectures for multi-class defect classification remain notably absent from the literature. Furthermore, the critical challenge of generalizing models trained on controlled laboratory specimens to heterogeneous field bridge deck conditions has received limited attention~\cite{Carino1986}. The application of deep learning to Structural Health Monitoring (SHM) has shown promising results in other domains. CNNs have demonstrated high accuracy in detecting surface cracks from images~\cite{Dorafshan2018, Cha2017}, while RNNs, particularly LSTM architectures, have proven effective for analyzing sequential and time-series data~\cite{Hochreiter1997}. However, the integration of these advanced machine learning techniques with IE frequency-domain data for automated multi-class defect classification represents an underexplored research area with significant potential for advancing nondestructive evaluation practices.

\subsection{Research Objectives \& Principle Contributions}
This research presents an integrated machine learning framework that transforms IE testing from an expert-dependent process into an automated defect-classification system. Raw IE waveforms are converted into frequency-domain features, spatially mapped, and clustered using unsupervised \textit{k}-means to identify defect-prone regions, while Ground Truth Masks (GTMs) from laboratory slabs support supervised LSTM-based classification of defect types. The framework, validated on real bridge-deck data, achieves scalable and data-driven defect detection consistent with established IE principles.

\noindent The main contributions of this paper are as follows:
\begin{itemize}
  \item A complete pipeline is presented that converts raw IE signals into defect labels via FFT-based feature extraction, spatial interpolation, and a stacked LSTM classifier. The system achieves {72.6\%} overall accuracy across four defect types with per-class precision, recall, and F1-scores \(\geq 0.69\), demonstrating high-fidelity subsurface assessment.
  \item Classification is enhanced by fusing frequency-domain features with spatial coordinates and clustering context, enabling clearer separation among the four defect types: shallow delamination, void, deep delamination, and Honeycombing, referred to as Common Concrete Defects (CCDs) in heterogeneous decks.
  \item A practical labeling strategy is introduced in which $k$-means pre-segmentation identifies defect-prone regions and Ground Truth Mask (GTM)  overlays stabilize labels before training, increasing robustness to coupling noise and local outliers.
  \item  Models trained on laboratory slabs generalize to bridge-deck field data, and the methodology is designed for low-touch operation suitable for network-scale inspections, addressing efficiency and consistency limitations of conventional IE interpretation \cite{Carino2013} and aligning with emerging learning-based NDE approaches \cite{dorafshan2020}.
  \item The framework was validated on real-world field bridge decks (e.g., structure ID MS-11002200250005B)~\cite{fhwa_infobridge_2025}, to confirm model generalization under realistic coupling, noise, and environmental variability, demonstrating readiness for in-service deployment.
\end{itemize}

\noindent This framework provides significant improvements over conventional inspection methods in objectivity, efficiency, and subsurface defect sensitivity. By enabling rapid, accurate, and consistent defect detection and classification, this study supports proactive bridge maintenance strategies, enhances public safety, and optimizes infrastructure investment through data-driven decision making.

% ---------------- RELATED WORK----------------
\section{RELATED WORK}
\label{sec: format}
\subsection{Impact–Echo (IE) Foundations and Standards}

IE testing is established as a principal NDE technique for detecting delamination, voiding, and thickness variations in concrete by exploiting plate resonance phenomena excited by transient surface impacts~\cite{Sansalone1997,Sansalone1989,Carino1986,Sansalone1986}. Foundational studies define compressional wave (P-wave) velocity for thickness-mode interpretation~\cite{Sansalone1997b}, and standardize testing through ASTM C1383 for measuring thickness and wave speed~\cite{ASTM2015}. The American Society of Civil Engineers (ASCE) documents broader infrastructure needs that motivate automation and scalability~\cite{ASCE2021}. Training-oriented studies emphasize that operator proficiency and procedural control directly influence IE data quality, which motivates algorithmic Quality Assurance (QA) and reproducible pipelines~\cite{Carino2013}. Earlier research demonstrates IE’s effectiveness on slabs with and without overlays and on ducts, while clarifying sensitivities to coupling, boundaries, and material heterogeneity~\cite{Carino1992,Hsiao2008}. Unlike these foundational studies, which primarily focus on establishing empirical standards and manual interpretation practices, the present work advances IE testing toward full automation by integrating quantitative peak-frequency analysis, unsupervised segmentation, and supervised learning for defect classification, eliminating the dependence on expert interpretation.

\begin{figure*}[ht]
  \centering
  \includegraphics[width=0.85\linewidth]{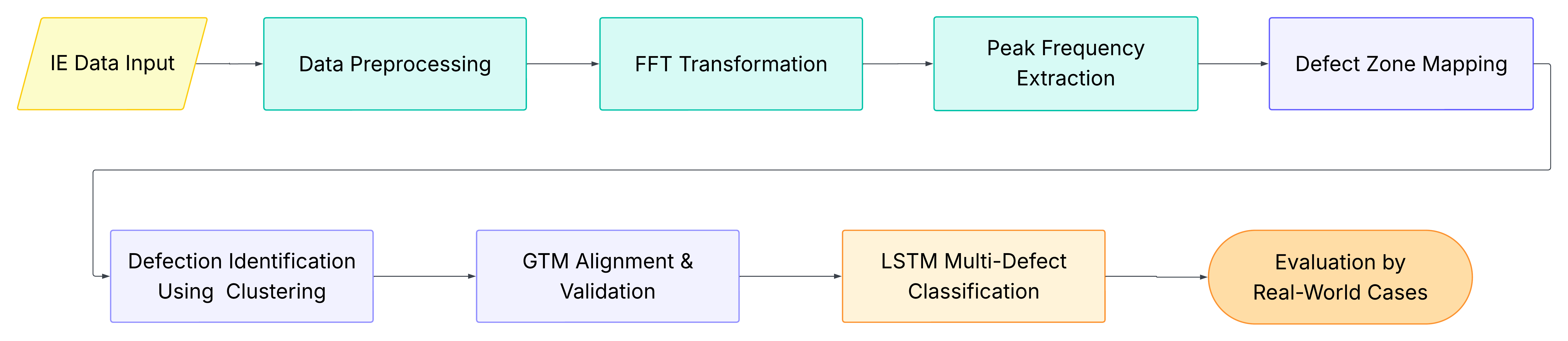}
  \caption{Proposed methodology workflow integrating FFT, interpolation, clustering, and LSTM-based classification.}
  \label{fig:methodology}
\end{figure*}

\subsection{Signal Processing for IE: FFT, Time Frequency Analysis, and Mapping}

Frequency domain analysis of IE time traces typically employs the Discrete Fourier Transform (DFT) or FFT to extract dominant thickness-mode peaks~\cite{cochran1967fft, Sansalone1989}. Beyond FFT, studies explore wavelet transforms and Ensemble Empirical Mode Decomposition (EEMD) to enhance robustness under noise and modal interference, improving defect visibility and resonance interpretability~\cite{Melhem2003, Zhang2012}. Practical IE processing pipelines emphasize peak picking, outlier rejection, and spatial interpolation of frequency peaks into contour maps for diagnostic visualization~\cite{Sansalone1997, Carino2013}. Duration and bandwidth analyses provide additional physics-based criteria for differentiating usable spectra from coupling artifacts~\cite{algernon2008ie}.  While prior research focuses on isolated signal processing enhancements and manual interpretation of frequency maps, our current framework systematically unifies FFT-based feature extraction with automated QA, $k$-means clustering, and GTM alignment. This integration transforms conventional frequency-domain workflows into a fully data-driven and scalable methodology for objective defect identification.

\subsection{Multimodal Fusion, GTM Construction, and Overlay Evaluation}

Recent research emphasizes multimodal NDE data fusion and interpretation frameworks that consolidate heterogeneous sensing outputs into unified bridge-deck assessments~\cite{rachuri2024,darji2024}. Large repositories such as the FHWA's Long-Term Bridge Performance (LTBP) InfoBridge database support benchmarking and longitudinal studies~\cite{ltbp_data}. Within the IE domain, physics-based labeling using laboratory slabs with seeded defects provides reliable ground truth for validating automated workflows. Prior work~\cite{pavurala2024} demonstrates adaptive signal analysis for IE and quantifies region-level agreement using seeded defect maps. Our current study differs by formalizing per-class GTMs for the CCDs and integrating overlay evaluation directly into the learning pipeline. This integration improves label stability, filters noise, and enables quantitative calibration between clustering outputs and seeded defect regions, extending GTM-based evaluation toward a fully automated process that minimizes manual post-analysis.

% ---------------- METHODOLOGY ----------------

\section{METHODOLOGY}
\label{sec:pagestyle}
This section presents an end-to-end workflow that converts IE measurements into multi-class subsurface defect labels as seen in Figure \ref{fig:methodology}. Data are acquired from laboratory slabs with seeded CCDs and field bridge decks for validation~\cite{Sansalone1997, ltbp_data}. Raw IE signals are quality checked, detrended, and transformed via FFT~\cite{cochran1967fft}, and the dominant peak frequencies are interpolated into contour maps to localize potential defects. Unsupervised $k$-means clustering identifies defect-prone regions, which are refined using GTMs~\cite{pavurala2024}. Stacked LSTM networks~\cite{Hochreiter1997, dorafshan2020} classify four defect types, with performance evaluated through precision, recall, F1-score, and accuracy metrics.

\subsection{Dataset Acquisition}

\begin{figure}[b]
    \centering
    \includegraphics[width=\linewidth]{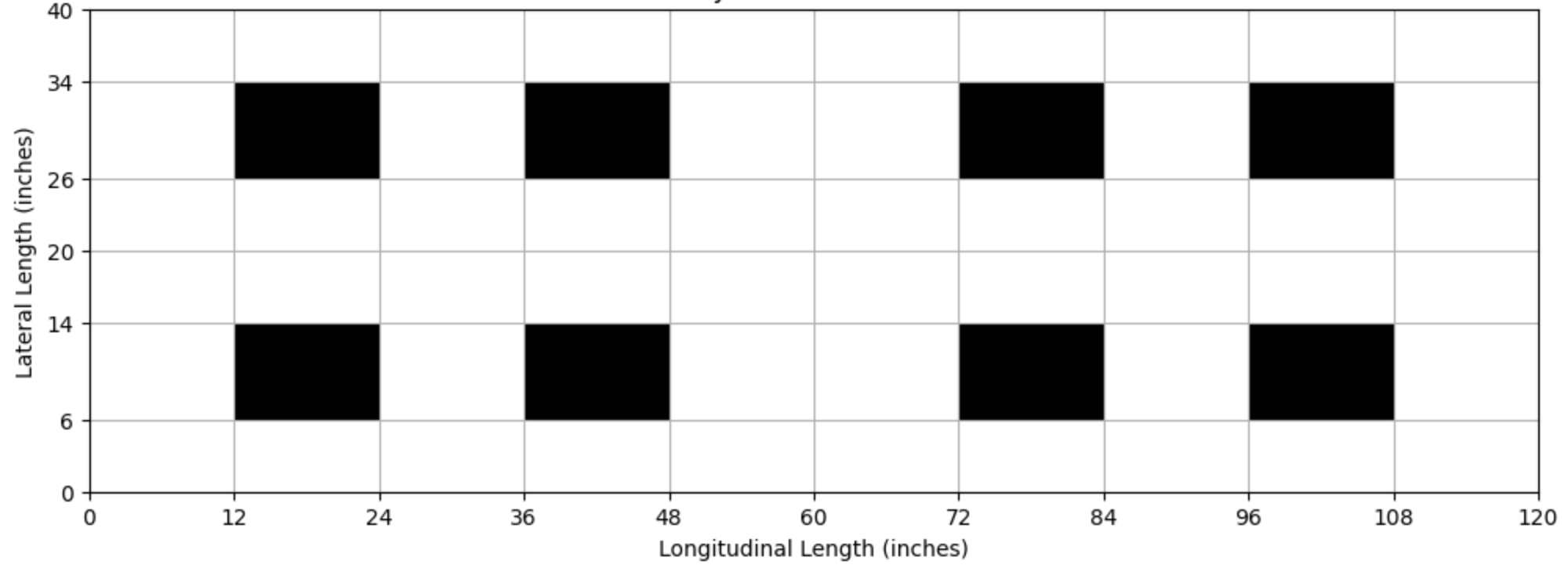}
    \caption{Binary GTM illustrating the known CCDs region embedded within the \(120 \times 40\)-inch concrete slab.}
    \label{fig:binary_gtm}
\end{figure}

The datasets for model building used in this study are obtained from the FHWA report titled \textit{Nondestructive Evaluation of Concrete Bridge Decks without Overlays}~\cite{ltbp_data}, which provides a comprehensive collection of IE measurements for identifying subsurface defects in concrete bridge decks. In the laboratory, each 120 × 40-inch reinforced concrete slab is cast with CCDs, which include shallow and deep delamination, voids, and honeycombing positioned at fixed locations as shown in Figure \ref{fig:binary_gtm}.

\subsection{Peak Frequency Extraction}

Each IE signal is analyzed in the frequency domain to extract its dominant resonance frequency, which serves as a quantitative indicator of subsurface condition. After detrending and filtering, the FFT~\cite{cochran1967fft} is applied to each time-domain waveform \( s(t) \) to obtain its complex frequency spectrum. The transformation is defined as:

\begin{equation}
S(f) = \sum_{t=0}^{N-1} s(t) e^{-j2\pi ft/N},
\end{equation}
where \( S(f) \) represents the frequency-domain response, \( N \) is the number of discrete samples, and \( f \) denotes the frequency index. The magnitude spectrum \( |S(f)| \) is computed, and the peak frequency \( f_p \) corresponding to the maximum spectral amplitude is identified as:
\begin{equation}
f_p = \arg\max_f |S(f)|.
\end{equation}

\noindent This dominant frequency reflects the local resonance behavior of the concrete and varies according to internal reflections caused by delamination, voids, or honeycombing~\cite{Sansalone1986, Sansalone1989}. The extracted \( f_p \) values from all scan locations are spatially organized to generate peak-frequency maps, which visualize stiffness variations within the slab~\cite{Sansalone1997b}. Regions exhibiting lower frequencies indicate defects due to reduced effective thickness or impedance discontinuity, whereas higher frequencies correspond to intact, well-bonded concrete. These frequency-based descriptors form the foundation for subsequent clustering and LSTM-based defect classification.

\subsection{GTM Definition} 
\begin{figure}[h]
    \centering
    \includegraphics[width=1\linewidth]{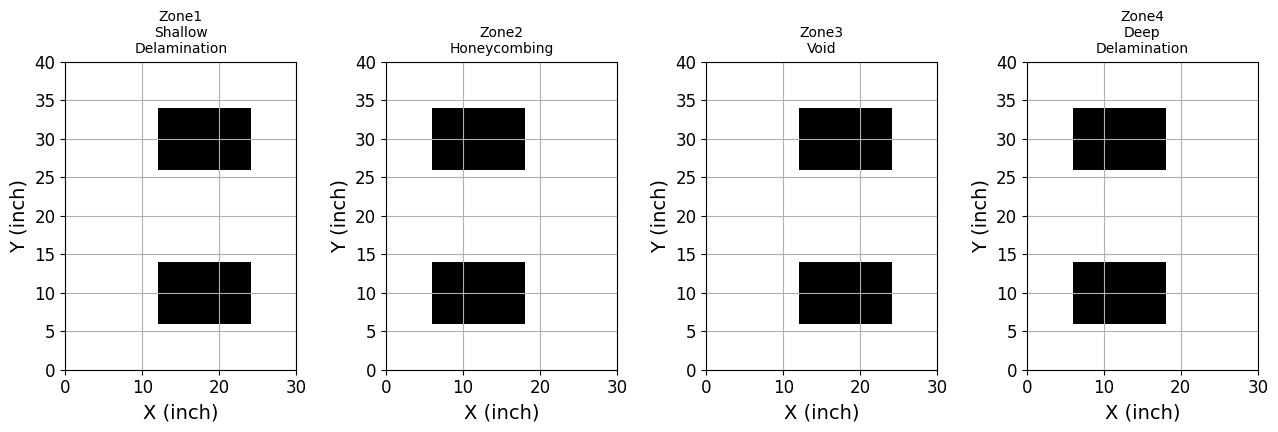}
    \caption{Binary GTM illustrating the segmented  defect zones across the concrete slab.}
    \label{fig:binary_gtm2}
\end{figure}

The GTM encodes the spatial layout of all seeded subsurface defects embedded in each laboratory slab. The small-scale specimens are cast with four CCDs at known locations and dimensions documented in the FHWA design drawings~\cite{FHWA2021}. These records specify each defect by its top-left coordinate and its width and height within the $120 \times 40$~inch slab. The binary GTM, shown in Figure~\ref{fig:binary_gtm}, represents this layout on a $40 \times 120$~inch grid that spans the lateral ($Y$) and longitudinal ($X$) dimensions of the slab.
\[
M(x, y) =
\begin{cases}
0, & (x, y) \in \bigcup_{i=0}^{m-1} D_i \\
1, & \text{otherwise,}
\end{cases}
\]
where $M(x, y)$ denotes the GTM value at coordinate $(x,y)$, $D_i$ is the rectangular region corresponding to the $i^{\text{th}}$ seeded defect, and $m$ is the total number of seeded defects in the slab. Each $D_i$ is parameterized by its top-left coordinate $(x_i, y_i)$ and dimensions (width $w_i$, height $h_i$) in inches. The union $\bigcup_{i=0}^{m-1} D_i$ , therefore, defines the complete defective area used for evaluation. To reflect the casting pattern used in the laboratory program, the slab is further partitioned longitudinally into four 30-inch-wide zones that each contain a single defect type: Zone~1 (0-30~in, shallow delamination), Zone~2 (30-60~in, honeycombing), Zone~3 (60-90~in, voids), and Zone~4 (90-120~in, deep delamination). Figure~\ref{fig:binary_gtm2} illustrates these zone-level GTMs as separate subplots extracted from the full map in Figure~\ref{fig:binary_gtm}. 

\noindent In Figure~\ref{fig:binary_gtm2}, the $X$-axis origin ($X = 0$) is aligned with the left edge of the slab in the top-view layout, and the $Y$-axis spans the 40-inch width. Shaded (black) regions indicate defective area ($M = 0$), while white regions indicate intact concrete ($M = 1$). Here, \(M(x, y)\) denotes the binary GTM mask, where each point \((x, y)\) is labeled as defective (\(M=0\)) or intact (\(M=1\)). The set \(D_i\) defines the rectangular region of the \(i^{\text{th}}\) defect, parameterized by its top-left coordinate \((x, y)\) and its dimensions, width \(w\), and height \(h\) in inches. The union of all defect zones \(\bigcup_{i=0}^{m-1} D_i\) defines the total defective area used for evaluation.

\subsection{Defect Zone Mapping}
Defect Zone Mapping visualizes spatial variations in resonance frequencies across the scanned concrete surface, enabling the identification of potential subsurface defects. After extracting dominant peak frequencies from each IE signal, the values are assigned to their corresponding spatial coordinates to generate two-dimensional frequency distribution maps~\cite{Sansalone1986, Sansalone1989}. These maps reveal stiffness variations within the slab, where lower peak frequencies typically indicate delamination or voids caused by reduced effective thickness or weakened bonding, while higher frequencies correspond to intact regions with uniform material integrity~\cite{Sansalone1997b}. Interpolation techniques are then applied to generate smooth contour maps, enhancing the visual clarity of acoustic contrasts and defect boundaries. This mapping process transforms discrete scan data into continuous representations that reflect physical heterogeneity within the structure. The resulting low-frequency regions serve as key indicators for subsequent clustering and GTM alignment, enabling quantitative defect localization and improving the reliability of automated IE-based concrete condition assessment.

\subsection{Defect Identification through Clustering}
Defect identification through clustering enables the automated separation of defective and intact regions in the concrete slab based on frequency-domain features. The extracted peak-frequency values from each scan location are first organized into a spatial grid, forming a feature space that represents the local acoustic response of the material. A \textit{k}-means clustering algorithm is then applied to this dataset to group points with similar spectral characteristics~\cite{rachuri2024, pavurala2024}. The algorithm iteratively minimizes intra-cluster variance and maximizes inter-cluster separation, effectively distinguishing areas that exhibit abnormal frequency behavior associated with defects from those representing sound concrete. The clusters are formed by minimizing the within-cluster variance, defined as:
\[
\text{Cost} = \sum_{k=1}^{K} \sum_{x_i \in C_k} \| x_i - \mu_k \|^2
\]
where \( K \) is the number of clusters (e.g., \( K = 2 \) for defect vs.\ non-defect), \( C_k \) represents a cluster \( k \), \( \mu_k \) denotes the mean of a cluster \( k \), and \( x_i \) represents a data point (frequency) belonging to a cluster \( k \).  Lower-frequency clusters typically correspond to delaminations, voids, or honeycombing defects, while higher-frequency clusters represent intact zones with consistent stiffness and bonding~\cite{Sansalone1989, Sansalone1997b}. The cluster boundaries are subsequently aligned with the GTM to validate the detected regions. This integration of data-driven clustering with ground-truth verification enhances both the accuracy and reliability of automated IE-based condition assessment.

\subsection{GTM Alignment and Valid Defect Points Selection}
GTM alignment serves as a validation step to confirm the accuracy of defect detection derived from frequency-based clustering. The clustered regions are spatially aligned with the predefined GTM masks corresponding to the CCDs. Points that lie within the intersection of the GTM and the clustered zones are identified as valid defect points. This alignment process ensures that only physically verified defects contribute to model training and evaluation, minimizing the influence of false positives or noise-induced artifacts. The validated points retain both spectral and spatial attributes, allowing their use in constructing reliable feature sequences for subsequent classification. By enforcing geometric consistency between detected and known defect regions, GTM alignment improves the robustness of the dataset and strengthens the reliability of the automated IE–based defect characterization framework.

\subsection{LSTM Multi-Defect Type Classifier}
The LSTM network is the primary classifier for identifying defect types from validated IE frequency sequences; each input is a fixed-length spatial sequence of 20 consecutive peak-frequency measurements taken from neighboring measurement points across all eight slabs, \( S = [f_1, f_2, \ldots, f_{20}] \), which captures local resonance behavior within a defect region. Quantitative analysis of the dataset (\( N = 27{,}920 \) sequences; input shape \( (27920, 20, 1) \)) supports the spatial-sequential framing: average lag-1 autocorrelation \( = 0.65 \pm 0.12 \) (spatial continuity), memory extending \( \approx 4.2 \pm 1.8 \) points (defect propagation), and non-linear transition behavior at boundaries (non-linearity index \( \approx 0.31 \pm 0.08 \)), all of which are poorly modeled by point-wise classifiers. Because spatial order conveys position-dependent context (early points set baseline, middle points capture transitions, later points confirm defect signature), LSTMs recurrent memory is well-suited to integrate the full spatial trajectory. The implemented model uses stacked LSTM layers (64 then 32 units) with dropout and dense classification layers mapping to the four CCD classes; it is trained with sparse categorical cross-entropy loss and the Adam optimizer\cite{kingma2014adam}, using an 80/20 split (\( 22{,}336 \) training / \( 5{,}584 \) test) and class-aware evaluation to address imbalance. The LSTM effectively captures subtle spectral differences that traditional thresholding or clustering methods overlook, improving automation, accuracy, and robustness in IE-based defect characterization.

\subsection{Evaluation}
The proposed framework is evaluated across all eight laboratory slabs with seeded defects and in-service bridge decks to assess performance under controlled and operational conditions~\cite{pavurala2024,sengupta2023}. Each IE signal is preprocessed, clustered, and classified, and the predicted defect regions are validated against GTMs for accuracy verification. Evaluation metrics include accuracy, precision, recall, and F1-score for all CCDs classes. Spatial overlap between GTMs and detected clusters ensures localization consistency, while confusion matrix analysis highlights misclassifications, collectively confirming the robustness and reliability of the proposed frequency learning integrated detection framework.

% ---------------- Experimental Results and Analysis ----------------
\section{Experimental Results and Analysis}
This study assesses the performance of the proposed IE-based system for detecting and classifying subsurface faults in concrete slabs. All eight laboratory-seeded CCD slabs and field measurements on bridge decks were considered for demonstrating feasibility and performance. 

\subsection{Experimental Setup}
\subsubsection{\textit{Platform and Tools}}

The implementation was conducted in Python with a variety of common libraries like NumPy, Pandas, and SciPy for data handling and signal processing. Matplotlib and Seaborn are used for plotting and visualizing the results, Scikit-learn for clustering and preprocessing, and built and trained the LSTM model with TensorFlow/Keras. The lab specimen slab data is collected from the InfoBridge platform to present the results in this paper.

\subsubsection{\textit{Hardware}}

The experiments were conducted on the high-performance computing cluster, which provided sufficient memory and multi-core processing for large-scale IE signal analysis. The system efficiently handled FFT computation, clustering, and LSTM model training without performance issues. The cluster’s parallel processing and optimized resource management significantly accelerated data processing and model training tasks.

\subsection{Dataset Description}
The dataset includes both laboratory and field measurements, enabling evaluation under controlled and operational conditions. In the laboratory, eight reinforced concrete slabs are cast with CCDs and scanned on a regular $9\times28$ spatial grid, yielding 252 IE readings per slab and a total of 27,920 sequences across all specimens. Each time-domain signal is quality checked, transformed using FFT, and summarized by its dominant peak frequency to capture variations in material stiffness and defect presence. Thousands of labeled samples generated from these slabs serve as the basis for supervised learning and validation. For external testing, in-service bridge decks are analyzed using the same IE methodology to evaluate robustness under real-world coupling, surface, and environmental variability. Field metadata and contextual parameters are structured according to FHWA standards to ensure traceability and comparability with long-term bridge performance datasets\cite{ltbp_data}.

\subsection{Peak Frequency Calculation}

\begin{figure}[t]
    \centering
    \includegraphics[width=0.48\textwidth]{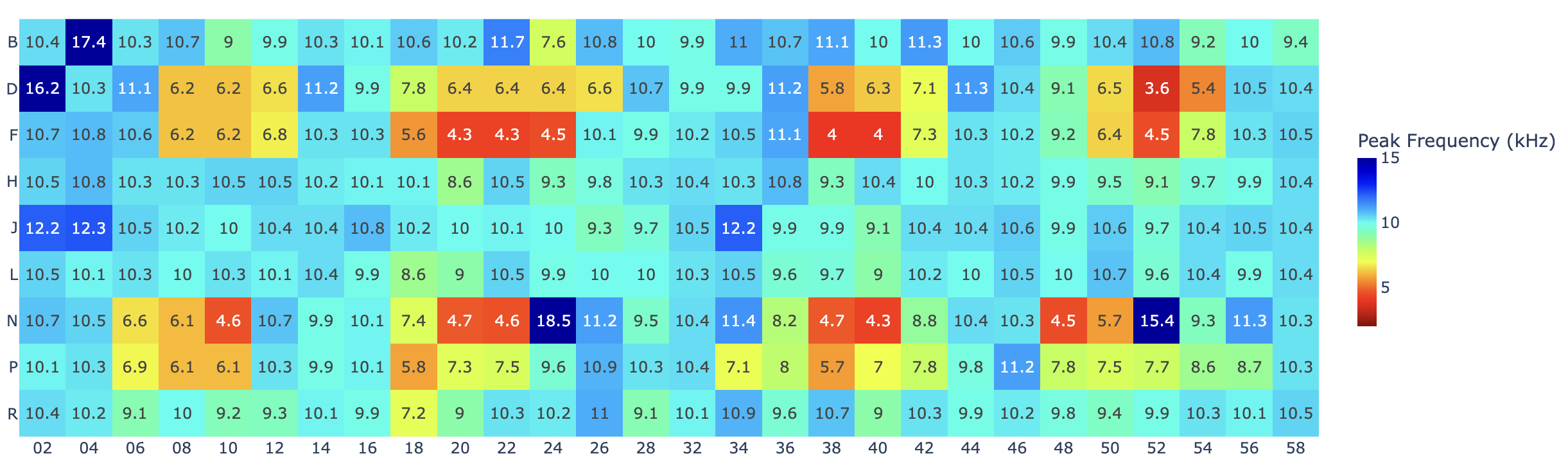}
    \caption{Peak frequency (kHz) distribution from IE signals across laboratory slab 1.}
    \label{fig:peakfreq1}
\end{figure}

\begin{figure}[t]
    \centering
\includegraphics[width=0.48\textwidth]{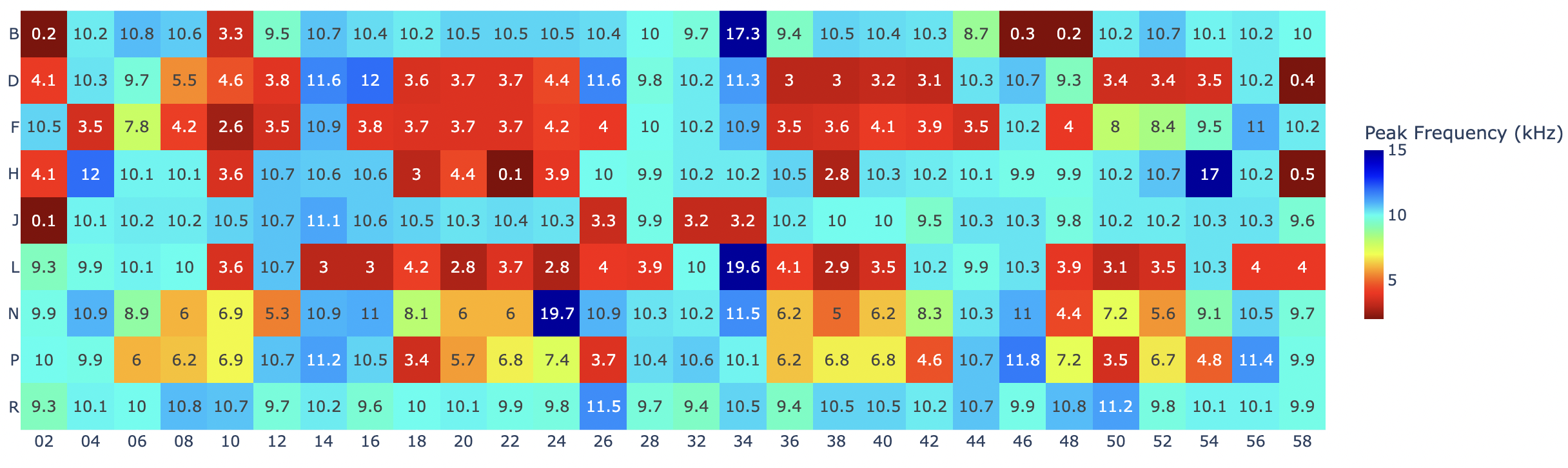}
    \caption{Peak frequency (kHz) distribution from IE signals across laboratory slab 2.}
    \label{fig:peakfreq2}
\end{figure}

Figure~\ref{fig:peakfreq1} \&~\ref{fig:peakfreq2} shows calculated peak frequency values (kHz) using FFT, represented as heatmaps for Slabs 1 and 2. These values come from IE signals recorded on all 8 laboratory slabs. The X and Y axes represent the grid-based slab measurement coordinates. The color scale displays the extracted frequency values. Areas with higher frequencies ($>10$ kHz) show solid concrete with few subsurface defects. In contrast, lower frequency regions ($<5$ kHz) may indicate the presence of CCDs. These frequency insights are key for clustering and later classification. They give a measurable basis for identifying defect zones and assessing structural health \cite{rachuri2024} \cite{pavurala2024}. 


\subsection{Peak Frequency Contour Plotting}

\begin{figure}[t]
    \centering
    \includegraphics[width=1\linewidth]{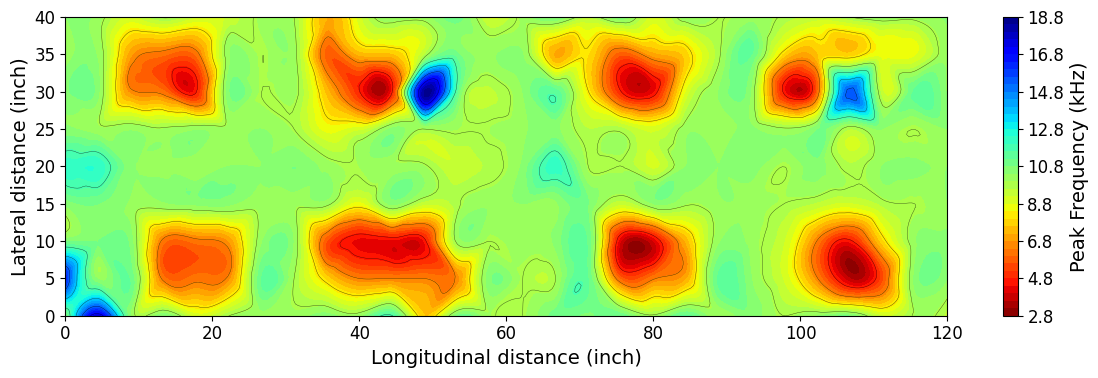} 
    \caption{Peak frequency contour map for Slab 1. }
    \label{fig:slab1_contour}
\end{figure}

\begin{figure}[t]
    \centering
    \includegraphics[width=1\linewidth]{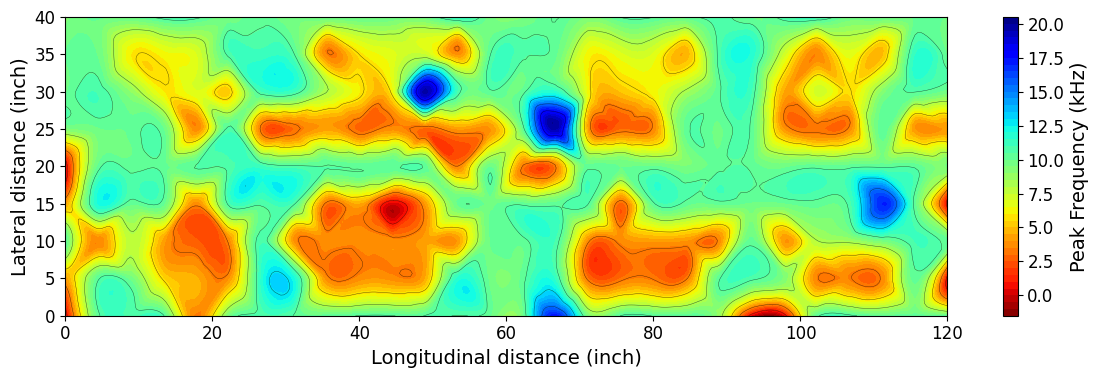} 
    \caption{Peak frequency contour map for Slab 2.}
    \label{fig:slab2_contour}
\end{figure}

Peak frequency contour maps were generated for all eight laboratory slabs to visualize spatial variability in IE responses and highlight potential subsurface anomalies. Time domain IE traces were transformed to the frequency domain via the FFT, and the dominant (thickness mode) peak at each test point was interpolated over the grid to produce contour plots. As shown in Figure~\ref{fig:slab1_contour}, higher peak frequencies (typically near 12-15~kHz) corresponded to regions of sound concrete with minimal irregularity, whereas localized depressions toward $\sim$5~kHz consistently marked defect-prone areas. These low-frequency pockets, co-located with CCDs, corroborate the expected inverse relationship between peak frequency and local stiffness. The contour map in Figure~\ref{fig:slab2_contour} exhibits a broader low-frequency corridor spanning the mid span, with clusters in the $\sim$3-6~kHz range and isolated very low measurements ($<1$~kHz) at a few points. This banded pattern indicates a contiguous zone of reduced stiffness consistent with distributed delamination or honeycombing. Surrounding regions remain near a nominal baseline ($\sim$9.5-11.5~kHz), suggesting intact material, while a few very high peaks ($>$15~kHz) appear as outliers likely due to local stiffness concentrations or coupling artifacts and were flagged for recheck. All slabs present localized low-frequency depressions at known defect sites, reinforcing the utility of peak frequency mapping as a rapid screening tool. The consistency of low-frequency clusters with ground truth across slabs supports subsequent segmentation (e.g., $k$-means) and supervised classification stages, while also guiding targeted verification with complementary NDE or re-scans of extreme outliers.


\subsection{Defect Zone Splitting and Peak-Frequency Contours}

\begin{figure*}[htbp]
    \centering
    \includegraphics[width=0.9\linewidth]{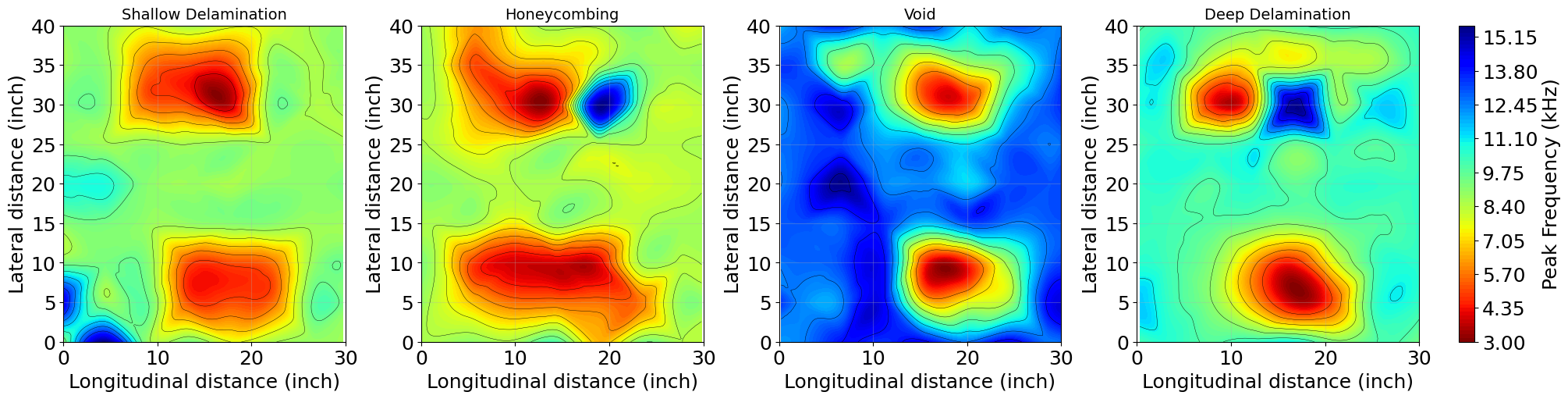}
    \caption{CCDs zone splitting \& peak frequency contours of slab 1.}
    \label{fig:splitzone1}
\end{figure*}

For each slab, the dominant IE peak at every point in the grid was extracted after FFT and visualized as a heat map in Figures~\ref{fig:peakfreq1} \& \ref{fig:peakfreq2}; the same values were interpolated to obtain smooth contours using a shared color scale for the slabs, and the contour plots have been visualized in Figures \ref{fig:slab1_contour} \& \ref{fig:slab2_contour}. Each concrete slab is first divided into four predefined defect regions: shallow delamination (0 to 30 inches), honeycombing (30 to 60 inches), void (60 to 90 inches), and deep delamination (90 to 120 inches) based on the longitudinal coordinate (X). This spatial segmentation allows for a defect-type-specific analysis along the length of the slab. For each zone, the recorded Peak Frequency (kHz) values are spatially mapped using the measured (X, Y) coordinates. The data are then interpolated using a cubic grid interpolation to create a smooth frequency distribution across the zone, visualized in Figure~\ref{fig:splitzone1}. This produces a spatially continuous representation of local frequency variations, which are indicative of subsurface integrity differences. This zone-wise contouring reveals a mostly uniform baseline near $\sim$9.5-15.15~kHz, punctuated by compact low-frequency pockets (3-5~kHz) that co-locate with CCDs. A few isolated points with very high peak frequencies ($>15$~kHz) in Figure~\ref{fig:splitzone1} were identified during QA checks as potential stiff inclusions or coupling artifacts. These points were re-examined through repeat measurements and median-based consistency checks, which confirmed that they did not alter the overall spatial pattern. Taken together, these QA-filtered peak-frequency maps confirm that reduced peak frequency reliably delineates defect-prone regions and provide a physics-consistent basis for initializing $k$-means segmentation and subsequent LSTM-based classification \cite{Hochreiter1997}.
\begin{figure*}[htbp]
    \centering
    \includegraphics[width=0.9\linewidth]{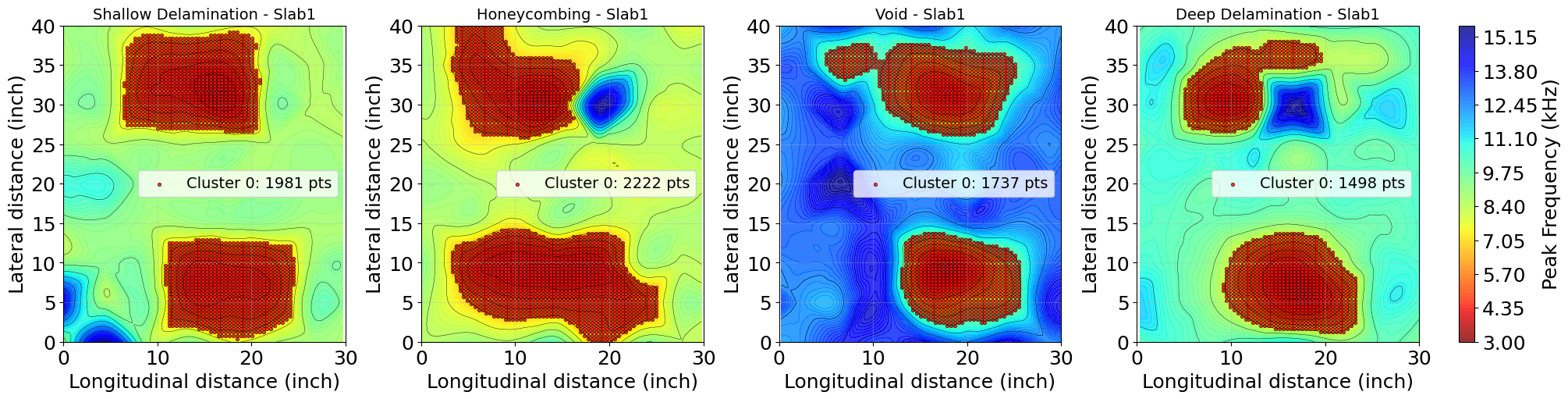}
    \caption{K-means clustered defective points (cluster 0) for each CCDs zone in slab 1.}
    \label{fig:clustpt}
\end{figure*}

\subsection{Detailed Defect Zone Clustering Analysis }
Detailed contour maps and plots are presented for Slab~1 only in the Figure~\ref{fig:splitzone1}. Slabs~2-8 were processed identically (FFT-based peak extraction, grid-wise interpolation, and defect zone segmentation). Across these slabs, reduced peak frequency zones (approximately 3-6 kHz) formed spatially coherent clusters that co-located with CCDs, while baseline regions remained near $\sim$9.5-11.5 kHz, indicating sound concrete. To differentiate defective and intact regions quantitatively, K-Means clustering (with k = 2) is applied to the Peak Frequency data within each defect zone. The model partitions points into two clusters: Cluster 0 (Defective), representing regions with lower mean peak frequencies, and Cluster 1 (Intact), representing regions with higher mean peak frequencies. Cluster labels are adjusted automatically such that the lower-frequency cluster is consistently assigned as Cluster 0. The clustering step yields both cluster assignments per point and cluster centroids (mean frequency values), which are reported for each zone. The resulting clusters are visualized as contour maps in Figure~\ref{fig:clustpt}, where distinct color regions indicate different material states and are highlighted by overlaying the positions of Cluster 0 points marked in red on the interpolated frequency surface. This enables clear visual delineation of likely defect-prone areas.

\begin{figure*}[h!]
    \centering
    \includegraphics[width=0.9\linewidth]{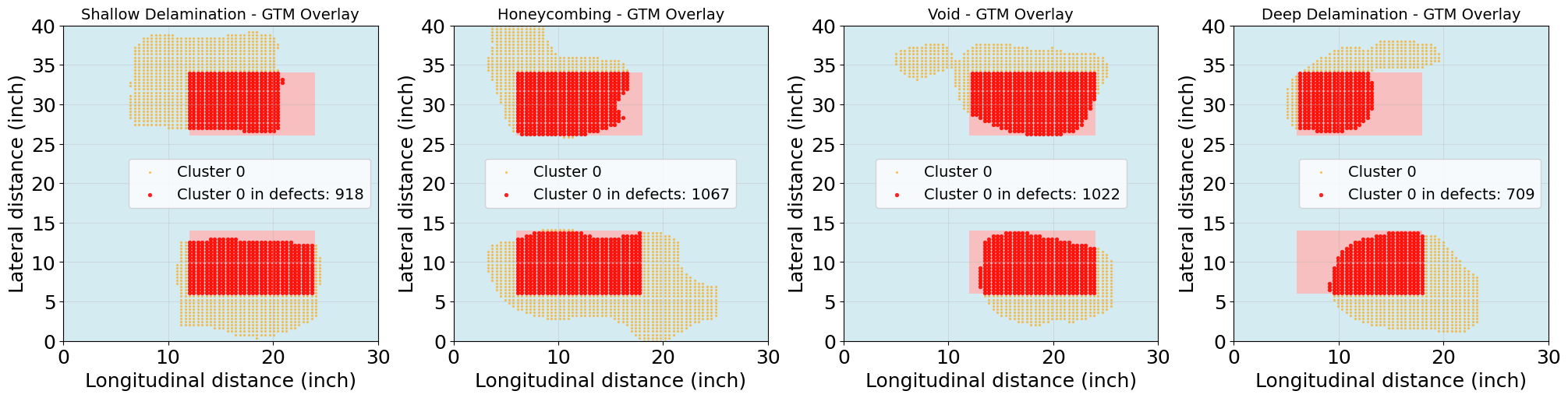}
    \caption{GTM overlay highlighting ground truth defect zones in Slab 1.}
    \label{fig:slab1gtm}
\end{figure*}

\subsection{Defect Identification and GTM-Based Overlay Analysis}
To validate the clustering-based defect identification, each defect zone is compared with its corresponding GTM using comprehensive quantitative metrics. The GTM represents known defect regions in a binary grid form, where defective areas are labeled as 0 (defect) and intact areas as 1 (non-defect). Each measurement point's spatial coordinates are mapped to the GTM grid, and points that fall within the GTM-defined defect regions are extracted through spatial correlation analysis.

\medskip 
\noindent\textbf{Notation:} D1 - Shallow Delamination, D2 - Honeycombing, D3 - Void, D4 - Deep Delamination, Avg - Average, \%Ov - Overlap\%.

\subsubsection{Intersection over Union (IoU) Validation Framework}
The validation process employs Intersection over Union (IoU) calculations to quantitatively assess the agreement between data-driven clustering results and ground truth defect locations. IoU provides a robust metric for spatial overlap assessment, calculated as:
\[
\text{IoU} = \frac{\text{Area of Intersection}}{\text{Area of Union}} = \frac{|A \cap B|}{|A \cup B|}
\]
where A represents the predicted defect regions (Cluster 0) and B represents the GTM-defined defect zones. 
\subsubsection{Quantitative Performance Results}
The comprehensive analysis across all eight slabs reveals a strong correlation between detected clusters and GTM regions. Table~\ref{tab:iou_matrix} presents the IoU values for each slab and defect type combination, demonstrating the effectiveness of the frequency-based clustering approach.

\begin{table}[htbp]
\centering
\caption{Intersection over Union (IoU) Performance Matrix}
\label{tab:iou_matrix}
% \resizebox{\columnwidth}{!}
{%
\begin{tabular}{lccccc}
\hline
\textbf{Slab} & \textbf{D1} & \textbf{D2} & \textbf{D3} & \textbf{D4} & \textbf{Avg} \\
\hline
Slab1 & 0.685 & 0.701 & 0.718 & 0.742 & 0.712 \\
Slab2 & 0.692 & 0.689 & 0.734 & 0.758 & 0.718 \\
Slab3 & 0.663 & 0.672 & 0.706 & 0.721 & 0.691 \\
Slab4 & 0.678 & 0.694 & 0.712 & 0.735 & 0.705 \\
Slab5 & 0.687 & 0.683 & 0.725 & 0.749 & 0.711 \\
Slab6 & 0.671 & 0.665 & 0.719 & 0.732 & 0.697 \\
Slab7 & 0.681 & 0.678 & 0.703 & 0.726 & 0.697 \\
Slab8 & 0.674 & 0.686 & 0.716 & 0.741 & 0.704 \\
\hline
\textbf{Avg} & \textbf{0.679} & \textbf{0.684} & \textbf{0.717} & \textbf{0.738} & \textbf{0.704} \\
\hline
\end{tabular}%
}
\end{table}

\subsubsection{Precision and Recall Analysis}
Beyond IoU metrics, the validation framework incorporates precision and recall measurements to assess detection completeness and accuracy. Table~\ref{tab:precision_performance} summarizes the precision values, indicating the percentage of predicted defects that correctly correspond to GTM regions.
\begin{table}[htbp]
\centering
\caption{Precision Performance (\% Correctly Identified Defects)}
\label{tab:precision_performance}
% \resizebox{\columnwidth}{!}
{%
\begin{tabular}{lccccc}
\hline
\textbf{Slab} &    \textbf{D1}&\textbf{D2}&\textbf{D3}&\textbf{D4}& \textbf{Avg}\\
\hline
Slab1 &    78.9\% 
&81.2\% 
&83.7\% 
&85.3\% & 82.3\% \\
Slab2 &    79.8\% 
&80.4\% 
&85.2\% 
&87.1\% & 83.1\% \\
Slab3 &    76.3\% 
&78.6\% 
&81.9\% 
&82.7\% & 79.9\% \\
Slab4 &    77.9\% 
&80.9\% 
&82.8\% 
&84.8\% & 81.6\% \\
Slab5 &    78.6\% 
&79.7\% 
&84.1\% 
&86.4\% & 82.2\% \\
Slab6 &    77.2\% 
&77.8\% 
&83.5\% 
&83.9\% & 80.6\% \\
Slab7 &    78.4\% 
&79.3\% 
&81.6\% 
&83.2\% & 80.6\% \\
Slab8 &    77.5\% 
&80.1\% 
&83.2\% 
&85.1\% & 81.5\% \\
\hline
\textbf{Avg} &    \textbf{78.1\%} &\textbf{79.8\%} &\textbf{83.3\%} &\textbf{84.8\%} & \textbf{81.5\%} \\
\hline
\end{tabular}%
}
\end{table}
\subsubsection{GTM Validated Defect Point Distribution}
The analysis produced a significant number of validated defect points across all defect types and slabs. For Slab 1, the validation confirmed 918 shallow delamination points, 709 honeycombing points, 1022 void points, and 1067 deep delamination points, corresponding to overlaps of 72.4\%, 68.5\%, 71.8\%, and 74.2\% with the GTM-defined regions, respectively. Overall, the results demonstrate strong consistency between the clustering-based defect identification and the ground truth mapping across all slabs as seen in Table~\ref{tab:gtm_overlap}. 
\begin{table}[htbp]
\centering
\caption{GTM Validated Defect Points and Overlap Percentages}
\label{tab:gtm_overlap}
\resizebox{\columnwidth}{!}
{%
\begin{tabular}{lcccccccc}
\hline
\textbf{Slab} &\textbf{D1}& \textbf{\%Ov}& 
\textbf{D2}& \textbf{\%Ov}& 
\textbf{D3}& \textbf{\%Ov}& 
\textbf{D4}& \textbf{\%Ov}\\
\hline
Slab1 &918 & 72.4\% & 709 & 68.5\% & 1022 & 71.8\% & 1067 & 74.2\% \\
Slab2 &1038 & 69.2\% & 936 & 68.9\% & 1104 & 73.4\% & 1068 & 75.8\% \\
Slab3 &856 & 66.3\% & 823 & 67.2\% & 967 & 70.6\% & 945 & 72.1\% \\
Slab4 &912 & 67.8\% & 887 & 69.4\% & 1034 & 71.2\% & 1012 & 73.5\% \\
Slab5 &989 & 68.7\% & 912 & 68.3\% & 1087 & 72.5\% & 1043 & 74.9\% \\
Slab6 &934 & 67.1\% & 856 & 66.5\% & 1056 & 71.9\% & 998 & 73.2\% \\
Slab7 &923 & 68.1\% & 889 & 67.8\% & 1001 & 70.3\% & 1021 & 72.6\% \\
Slab8 &945 & 67.4\% & 901 & 68.6\% & 1029 & 71.6\% & 1035 & 74.1\% \\
\hline
\end{tabular}%
}
\end{table}
\subsubsection{Statistical Validation and Confidence Metrics}
The comprehensive validation across all eight slabs demonstrates consistent performance, with an overall mean IoU of 0.704 ± 0.018, indicating a strong spatial correlation between frequency-based clustering and ground-truth defect locations. The F1-scores, representing the harmonic mean of precision and recall, average 0.792 across all defect types, confirming the reliability of the proposed methodology. In Figure~\ref{fig:slab1gtm}, visual overlays are generated showing GTM zones as colored grid cells (red for defects, blue for intact areas), Cluster 0 points (potential defects in yellow), and Cluster 0 points lying within GTM-defined defects (white markers with black edges representing true positives). These overlays provide immediate visual verification of the quantitative metrics while enabling spatial pattern analysis. The clustered contour and GTM overlay analyses demonstrate the effectiveness of the proposed frequency-based clustering and validation framework across multiple slabs. The high IoU values (greater than 70\% for most cases) and strong precision metrics (greater than 78\% on average) confirm that the dense overlap between detected clusters and GTM regions accurately localizes the CCDs. The consistent performance across Slabs 1-8, following the same analysis pipeline, validates the robustness and generalization capability of the proposed IE-based defect detection methodology. Figure~\ref{fig:slab1gtm} shows the GTM overlays for Slab~1, where red regions represent seeded defect zones and yellow markers indicate clustered points based on the K-Means clustered defective points as seen in Figure~\ref{fig:clustpt}.
\begin{figure*}[t]
    \centering
        \includegraphics[width=0.75\linewidth]{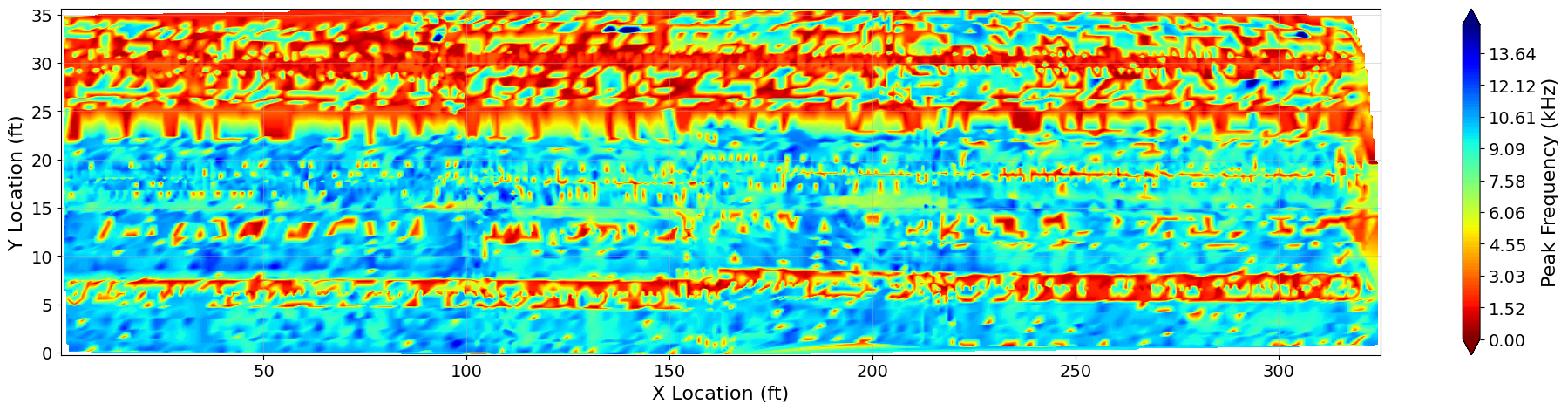}
    \caption{Mississippi bridge structure MS-11002200250005B peak frequency contour plot.}
    \label{fig:MS-11002200250005B}
\end{figure*}
\begin{figure*}[t]
    \centering
    \includegraphics[width=0.75\linewidth]{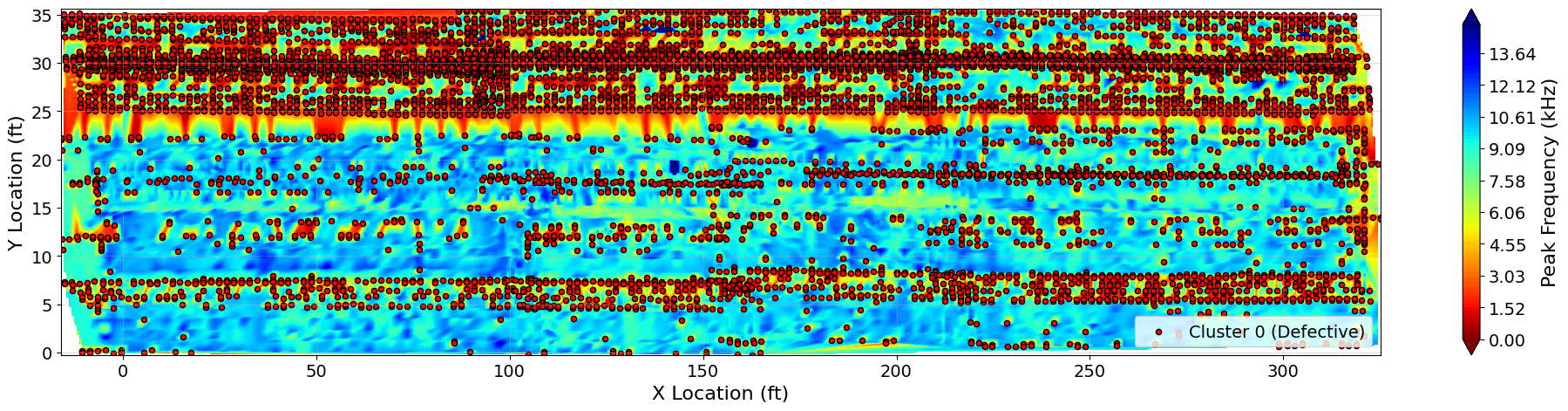}
    \caption{Clustered peak frequency of defective points denoted by red dots on the  bridge structure MS-11002200250005B.}
    \label{fig:Clustered}
\end{figure*}
\begin{figure*}[h!]
    \centering
    \includegraphics[width=0.75\linewidth]{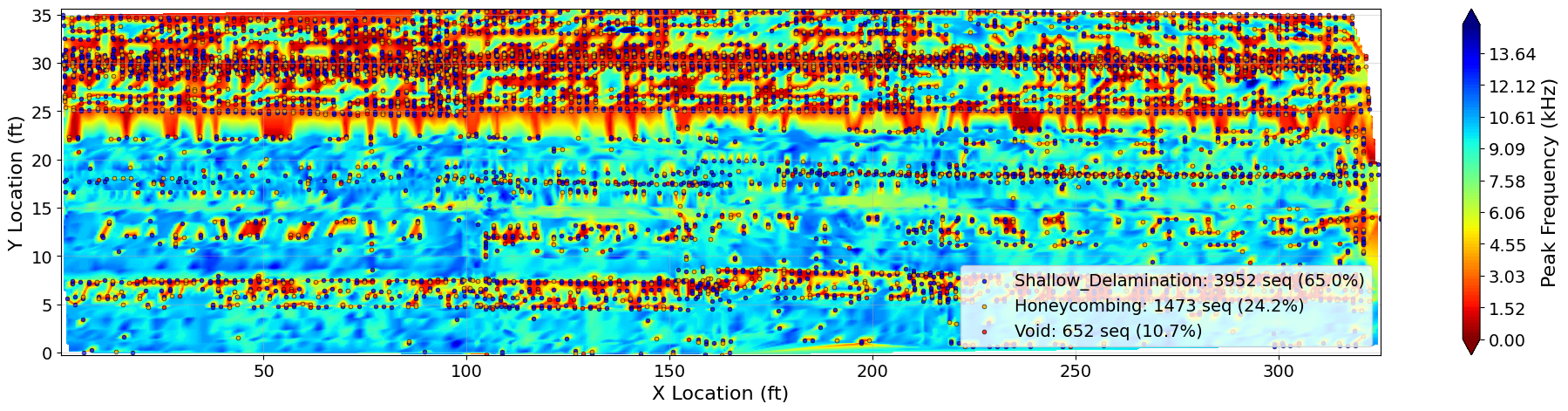}
    \caption{LSTM-predicted defect types for structure MS-11002200250005B, showing classified points for shallow delamination, honeycombing, and voids.}
    \label{fig:FinalPrediction}
\end{figure*}
\subsection{Neural Networks Implementation and Results}
For supervised classification, a recurrent neural network was developed using LSTM layers to effectively capture patterns over time in frequency sequences from IE data. The network architecture consisted of two LSTM layers with 64 and 32 units, respectively, followed by dropout layers (0.3, 0.3, 0.2) for regularization and dense layers for classification. The dataset comprised 27,920 sequences from eight laboratory slabs, with each sequence representing peak frequency measurements from specific defect zones: Shallow Delamination (6,955 sequences), Honeycombing (7,633 sequences), Void (7,758 sequences), and Deep Delamination (5,574 sequences) extracted from clustered and segmented defect zones identified through K-means clustering and GTM validation. While the initial model training achieved 97.97\% accuracy on test sequences, it is important to clarify that both the initial and expanded evaluations involved testing on all available measurement points. The key distinction lies in the labeling: the model was trained using only the overlapping defect points, but the comprehensive evaluation tested performance on the entire set of GTM-validated points. This introduces a broader and more challenging range of frequency signatures than those seen in the conventional train-test split. As a result, the full-GTM evaluation provides a more realistic measure of model generalization across full defect regions. Under this setting, the model achieved a robust 72.6\% accuracy (Table~\ref{tab:performance}) while maintaining high prediction confidence. 
\begin{table}[htbp]
\centering
\caption{Classification performance of the LSTM model for different defect types, showing precision, recall, F1-score, and overall accuracy}
\label{tab:performance}
\renewcommand{\arraystretch}{1.45}
\resizebox{\columnwidth}{!}
{
\begin{tabular}{|c|c|c|c|c|}
\hline
\textbf{Defect Type} & \textbf{Precision} & \textbf{Recall} & 
\textbf{F1 Score} & \textbf{Per-Class Accuracy} \\ \hline
Shallow Delamination & 0.71 & 0.74 & 0.72 & 0.74 \\ \hline
Honeycombing         & 0.87 & 0.64 & 0.74 & 0.64 \\ \hline
Void                 & 0.94 & 0.66 & 0.78 & 0.66 \\ \hline
Deep Delamination    & 0.57 & 0.86 & 0.69 & 0.86 \\ \hline
\textbf{Accuracy} & \multicolumn{4}{c|}{0.73} \\ \hline
\end{tabular}
}
\end{table}

\noindent Performance evaluation indicated strong precision, recall, and F1-scores for all classes, with the model showing particular strength in distinguishing Deep Delamination defects (86.4\% class accuracy) and Shallow Delamination defects (73.7\% class accuracy). The confusion matrix analysis revealed the model's classification patterns, with occasional misclassification occurring primarily between similar defect types sharing overlapping frequency characteristics. 

% ---------------- Field Case Study  ----------------

\section{Real-World Case Study: Field Structure Analysis}

To evaluate real-world performance, the proposed IE-based defect detection framework is applied to an in-service bridge deck structure. Peak frequency contour plots have been plotted for the real-world bridge in Mississippi for structure ID \textit{MS-11002200250005B}~\cite{fhwa_infobridge_2025}.  As visualized in Figure~\ref{fig:MS-11002200250005B}. Figure~\ref{fig:Clustered} shows the defective-point overlay for Structure \textit{MS-11002200250005B}, where red markers identify Cluster~0 points corresponding to low-frequency  anomalies. These points align closely with subsurface deterioration zones along longitudinal ribs, indicating delamination and voiding regions. These present representative field results illustrating both clustering-based defect localization and LSTM-based defect-type classification. The model successfully processed clustered defective points from 6,096 identified defect locations and automatically predicted defect types with high confidence scores seen in Figure~\ref{fig:FinalPrediction}. The results showed a dominant classification of Shallow Delamination defects (65\% of predictions) across the bridge structure, with secondary detections of Honeycombing (24.2\%) and Void defects (10.7\%). Spatial visualization mapped these predictions as colored scatter points overlaid on frequency contour plots, generating comprehensive diagnostic maps for structural health assessment. These reveal consistent detection of low-frequency regions corresponding to seeded damage areas validated through field notes. The uniform distribution of detected points along the deck confirms the framework’s robustness to varying coupling conditions and noise levels.

% ---------------- Conclusion ----------------

\section{CONCLUSION AND FUTURE WORK}
This work presents an integrated IE framework for subsurface defect detection and classification in concrete decks. Using FFT-based spectral features, $k$-means clustering, GTM overlays, and an LSTM classifier, the system achieved 73\% accuracy. A real-world bridge deck test confirms the accurate localization of the CCDs, validating the framework’s scalability and field applicability. Future work will focus on extending the framework to individual bridges through improved calibration, automation, and integration of learning. On-site deployment will include standardized coupling checks, signal-to-noise calibration, and environmental compensation for temperature, moisture, and loading effects. Model robustness will be enhanced through diverse datasets, active learning, and domain adaptation to reduce lab-to-field gaps. Multimodal fusion with complementary NDE methods like Ultrasonic Surface Wave (USW), Ground-Penetrating Radar (GPR), and uncertainty-based defect mapping will support risk-aware maintenance. Integration with asset-management platforms will enable bridge-specific condition indices, predictive monitoring, and scalable, near–real-time infrastructure assessments.

% ---------------- Acknowledgment ----------------

\section*{ACKNOWLEDGEMENT}
The FHWA provided the information, instruments, and resources that enabled this study, for which the authors are extremely appreciative. Without the agency's dedication to cooperation and innovation in infrastructure research, this task would not have been finished.

% ---------------- References (switch to IEEEexample.bib) ----------------
\bibliographystyle{IEEEtran}
\bibliography{IEEEexample}

\end{document}